\documentclass{amsart}
\usepackage{xspace}
\usepackage[usenames,dvipsnames]{pstricks}
\usepackage{epsfig}
\usepackage{graphicx}
\usepackage[normalem]{ulem}
\newtheorem{thm}{Theorem}
\newtheorem{cor}[thm]{Corollary}
\newtheorem{conj}[thm]{Conjecture}
\newtheorem{lem}[thm]{Lemma}

\theoremstyle{definition}

\theoremstyle{remark}

\newenvironment{prf}{{\bf \noindent Proof. } }{\hfill$\square$\\}

\newcommand{\ignore}[1]{}
\newcommand{\Wh}{well-hooped\xspace}
\newcommand{\wh}{well-hooped\xspace}

\newcommand{\extB}{buoy\xspace}
\newcommand{\extBs}{buoys\xspace}

\newcommand{\Enleve}[1]{{\xout{#1}} }
\begin{document}

\title[]{Reed's conjecture on some special classes of graphs}
\author{J.L. Fouquet, J.M. Vanherpe }
\address{L.I.F.O., Facult\'e des Sciences, B.P. 6759 \\
Universit\'e d'Orl\'eans, 45067 Orl\'eans Cedex 2, FR}
\subjclass{035 C} 

\begin{abstract}

Reed  conjectured that for any graph $G$, $\chi(G) \leq \lceil \frac{\omega(G)+\Delta(G)+1}{2}\rceil$, 
where $\chi(G)$, $\omega(G)$, and $\Delta(G)$ respectively denote the chromatic number, the clique number and the maximum degree of $G$. 
In this paper, we verify this conjecture for some special classes of graphs, in particular for subclasses of $P_5$-free graphs or $Chair$-free graphs.
\end{abstract}

\maketitle
\begin{center}
\today
\end{center}

\section{Introduction}
We consider here simple and undirected graphs. For terms which are not defined we refer to Bondy and Murty \cite{BonMur08}.

In 1998, Reed proposed the following Conjecture which gives, for any graph $G$, an upper bound of the chromatic number $\chi(G)$ in terms of the clique number $\omega(G)$ and the maximum degree $\Delta(G)$.

\begin{conj}[Reed's Conjecture \cite{Ree1998}] \label{Conjecture:Reed1998} For any graph $G$, $\chi(G) \leq \lceil \frac{\omega(G)+\Delta(G)+1}{2}\rceil$.
\end{conj}

\begin{figure}[t]
\scalebox{1} 
{
\begin{pspicture}(0,-0.83640623)(9.785937,0.79640627)
\psdots[dotsize=0.1](0.386875,0.7264063)
\psdots[dotsize=0.1](0.386875,0.22640625)
\psdots[dotsize=0.1](0.386875,-0.27359375)
\psdots[dotsize=0.1](0.886875,0.22640625)
\psdots[dotsize=0.1](0.886875,-0.27359375)
\psline[linewidth=0.03cm](0.886875,0.22640625)(0.886875,-0.27359375)
\psline[linewidth=0.03cm](0.386875,0.22640625)(0.886875,0.22640625)
\psline[linewidth=0.03cm](0.386875,0.7264063)(0.386875,-0.27359375)
\psdots[dotsize=0.1](1.986875,-0.27359375)
\psdots[dotsize=0.1](1.986875,0.22640625)
\psdots[dotsize=0.1](2.586875,-0.27359375)
\psdots[dotsize=0.1](2.586875,0.22640625)
\psdots[dotsize=0.1](2.286875,0.7264063)
\psline[linewidth=0.03cm](2.286875,0.7264063)(1.986875,0.22640625)
\psline[linewidth=0.03cm](2.586875,0.22640625)(2.586875,-0.27359375)
\psline[linewidth=0.03cm](1.986875,-0.27359375)(1.986875,0.22640625)
\psline[linewidth=0.03cm](1.986875,-0.27359375)(2.586875,-0.27359375)
\psline[linewidth=0.03cm](2.586875,0.22640625)(2.286875,0.7264063)
\psline[linewidth=0.03cm](1.986875,0.22640625)(2.586875,0.22640625)
\usefont{T1}{ptm}{b}{n}
\rput(0.57984376,-0.66859376){\small $Chair$}
\psdots[dotsize=0.1](3.786875,0.22640625)
\psdots[dotsize=0.1](4.086875,-0.27359375)
\psdots[dotsize=0.1](4.386875,0.22640625)
\psdots[dotsize=0.1](3.786875,0.7264063)
\psdots[dotsize=0.1](4.386875,0.7264063)
\psline[linewidth=0.03cm](3.786875,0.7264063)(3.786875,0.22640625)
\psline[linewidth=0.03cm](3.786875,0.22640625)(4.086875,-0.27359375)
\psline[linewidth=0.03cm](4.086875,-0.27359375)(4.386875,0.22640625)
\psline[linewidth=0.03cm](4.386875,0.22640625)(3.786875,0.22640625)
\psline[linewidth=0.03cm](4.386875,0.22640625)(4.386875,0.7264063)
\psdots[dotsize=0.1](5.886875,0.7264063)
\psdots[dotsize=0.1](5.586875,0.42640626)
\psdots[dotsize=0.1](6.186875,0.42640626)
\psdots[dotsize=0.1](5.886875,0.12640625)
\psdots[dotsize=0.1](5.886875,-0.27359375)
\psline[linewidth=0.03cm](5.886875,0.7264063)(6.186875,0.42640626)
\psline[linewidth=0.03cm](6.186875,0.42640626)(5.886875,0.12640625)
\psline[linewidth=0.03cm](5.886875,0.12640625)(5.886875,0.7264063)
\psline[linewidth=0.03cm](5.886875,0.7264063)(5.586875,0.42640626)
\psline[linewidth=0.03cm](5.586875,0.42640626)(5.886875,0.12640625)
\psline[linewidth=0.03cm](5.886875,0.12640625)(5.886875,-0.27359375)
\psdots[dotsize=0.1](7.586875,0.7264063)
\psdots[dotsize=0.1](7.286875,0.42640626)
\psdots[dotsize=0.1](7.886875,0.42640626)
\psdots[dotsize=0.1](7.586875,0.12640625)
\psdots[dotsize=0.1](7.586875,-0.27359375)
\psline[linewidth=0.03cm](7.586875,0.7264063)(7.886875,0.42640626)
\psline[linewidth=0.03cm](7.886875,0.42640626)(7.586875,0.12640625)
\psline[linewidth=0.03cm](7.586875,0.7264063)(7.286875,0.42640626)
\psline[linewidth=0.03cm](7.286875,0.42640626)(7.586875,0.12640625)
\psline[linewidth=0.03cm](7.586875,0.12640625)(7.586875,-0.27359375)
\psline[linewidth=0.03cm](7.286875,0.42640626)(7.886875,0.42640626)
\usefont{T1}{ptm}{b}{n}
\rput(2.3098438,-0.66859376){\small $House$}
\usefont{T1}{ptm}{b}{n}
\rput(3.9798439,-0.66859376){\small $Bull$}
\usefont{T1}{ptm}{b}{n}
\rput(5.7898436,-0.66859376){\small $Dart$}
\usefont{T1}{ptm}{b}{n}
\rput(7.469844,-0.66859376){\small $Kite$}
\psdots[dotsize=0.1](8.686875,0.32640624)
\psdots[dotsize=0.1](9.186875,0.5264062)
\psdots[dotsize=0.1](8.986875,0.02640625)
\psdots[dotsize=0.1](9.686875,0.32640624)
\psdots[dotsize=0.1](9.386875,0.02640625)
\psline[linewidth=0.03cm](9.186875,0.5264062)(9.686875,0.32640624)
\psline[linewidth=0.03cm](9.686875,0.32640624)(9.386875,0.02640625)
\psline[linewidth=0.03cm](9.186875,0.5264062)(8.986875,0.02640625)
\psline[linewidth=0.03cm](9.186875,0.5264062)(9.386875,0.02640625)
\psline[linewidth=0.03cm](9.386875,0.02640625)(8.986875,0.02640625)
\psline[linewidth=0.03cm](8.686875,0.32640624)(9.186875,0.5264062)
\psline[linewidth=0.03cm](8.686875,0.32640624)(8.986875,0.02640625)
\usefont{T1}{ptm}{b}{n}
\rput(9.209844,-0.66859376){\small $Gem$}
\end{pspicture} 
}
\label{fig:ChairHouseBullDartKite}
\caption{Configurations Chair, House, Bull, Dart,Kite}
\end{figure}
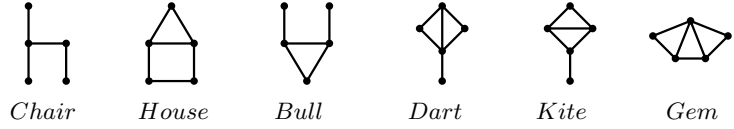

In \cite{AraKarSub2011}, Aravind et al. considered Conjecture \ref{Conjecture:Reed1998} for some graph classes defined by forbidden configurations.
In particular, when $P_n$, $C_n$ and $K_n$ respectively denote a chordless path, a chordless cycle and a complete graph on $n$ vertices while {\em Chair},
 {\em House}, {\em Bull}, {\em Dart} and {\em Kite} are the graphs depicted in 
Figure \ref{fig:ChairHouseBullDartKite}, Aravind et al. have shown that Conjecture \ref{Conjecture:Reed1998} holds for~:
\begin{itemize}
 \item ($P_5,\overline{P_2\cup P_3}, House, Dart$)-free graphs,
 \item ($P_5, Kite,Bull,(K_3\cup K_1)+K_1$)-free graphs,
 \item ($P_5, C_4$)-free graphs,
 \item ($Chair, House, Bull, K_1+C_4$)-free graphs,
 \item ($Chair, House, Bull, Dart$)-free graphs.
\end{itemize}


This paper proves that Reed's Conjecture holds for some classes of graphs. Our results extend those given in \cite{AraKarSub2011} on subclasses of $P_5$-free or $Chair$-free graphs.
\section{Notations and preliminary results}

\subsection{Odd hole expansions}

Given a graph $H$ on $n$ vertices $v_0\ldots v_{n-1}$ and a family of graphs $G_0\ldots G_{n-1}$, an {\em expansion} of $H$ (or {\em $H-$expansion}), denoted $H(G_0\ldots G_{n-1})$ is obtained from $H$ by replacing
each vertex $v_i$ of $H$ with $G_i$ for $i=0\ldots n-1$ and joining a vertex $x$ in $G_i$ to a vertex $y$ of $G_j$, ($i\neq j$) if and only if $v_i$ and $v_j$ are adjacent in $H$.
The graph $G_i$, $i=0\ldots n-1$ is said to be the {\em component} of the expansion associated to $v_i$.
For an expansion  $H(G_0\ldots G_{n-1})$ of some graph $H$, we will assume in the following that the vertices of $H$ are weighted with the chromatic number of their associated component
while an edge of $H$ is weighted with the sum of the weights of its endpoints.

 When $H$ is an odd hole, that is a chordless odd cycle of length at least $5$, we shall say that $G=H(G_0\ldots G_{n-1})$ is an {\em odd hole expansion}.

Conjecture \ref{Conjecture:Reed1998} was  studied by Rabern \cite{Rab2008}.

\begin{thm}\cite{Rab2008} \label{Theorem:Complement}If $\overline{G}$ is disconnected then
$\chi(G) \leq \lceil \frac{\omega(G)+\Delta(G) +1}{2}\rceil$.
\end{thm}

 Moreover~:

In \cite{AraKarSub2011} Aravind et al observed that the so-called {\em complete} expansion (every component of the expansion induces a complete graph) and {\em independent} expansion 
(every component of the expansion induces a stable) of an odd hole satisfy  Conjecture \ref{Conjecture:Reed1998}.
In \cite{FouVan2011} we have shown:

\begin{thm} \cite{FouVan2011} \label{Theorem:ReedsForBipartiteExpansion}
Any expansion of a bipartite graph satisfies  Conjecture \ref{Conjecture:Reed1998}.
\end{thm}

\begin{thm}\cite{FouVan2011} \label{Theorem:ChromaticNumberExpansionOddHole}
Let $G=H(G_0\ldots G_{2k})$ be an expansion of an odd hole $H$ of length $2k+1$ with $V(H)=\{v_0\ldots v_{2k}\}$and such that the edge $v_0v_1$ has maximum weigth in $H$. For $i=0\ldots {2k}$, 
let $\chi_i$ be the chromatic number of $G_i$.
Let $l$ be an index such that
$$\chi_{l-1}+\chi_l+\chi_{l+1}=\begin{matrix}\; \\ Min \\  _{3\leq i\leq 2k-1}\end{matrix} \left \{\chi_{i-1}+\chi_i+\chi_{i+1}\right \}.$$

Then
\begin{itemize}
 \item If $\chi_{0}+\chi_{1}\geq \chi_{l-1}+\chi_{l}+\chi_{l+1}$ then $\chi(G)=\chi_{0}+\chi_{1}$
 \item else $\chi(G)=\chi_{0}+\chi_{1}+\lfloor\frac{\chi_{l-1}+\chi_l+\chi_{l+1}-\chi_{0}-\chi_{1}+1}{2}\rfloor$.
\end{itemize}
\end{thm}

\begin{cor}\cite{FouVan2011} \label{Corollary:ReedsConjecturePourSpecialOddHole}
 Conjecture \ref{Conjecture:Reed1998} holds for an odd hole expansion when,
 in the conditions of Theorem \ref{Theorem:ChromaticNumberExpansionOddHole}, we have $\chi(A)=\omega(A)$ for $A\in \{G_{0},G_{1}, G_{l}\}$.
\end{cor}

\begin{thm} \cite{FouVan2011} \label{Theorem:C5_Expansion} If $G$ is a $C_{5}$-expansion then $G$ satisfies  Conjecture \ref{Conjecture:Reed1998}.
\end{thm}

\subsection{Notations and definitions}
$\;$\\ 
Let $X \subseteq V(G)$, $N(X)$ will denote the set of vertices in $V(G) - X$ adjacent to at least one vertex in $X$ while $G[X]$ will denote the subgraph of $G$ induced by $X$. 
If $X=\{v\}$ we write $G-v$ instead of $G[V(G)-X]$.
A vertex in $V(G)-X$ is said to be {\em partial} for $X$ if it is adjacent to some (but not all) vertex of $X$. 
As usual, given a graph $G$, $\omega(G)$, $\chi(G)$ and $\Delta(G)$ denote respectively the maximum number of vertices in a clique of $G$, the chromatic number and the maximum degree.
In addition, for a vertex $v\in V(G)$, $\omega(v)$ denotes the size of a maximum clique containing $v$, and $d(v)$ is the degree of $v$.

In \cite{FouGiaMaiThu1995}, a {\em buoy} was defined as a special case of $C_5$-expansion, that is an expansion of the odd hole $C_{5}$.
We extend here this notion to odd holes of length at least $5$. We shall say that an induced subgraph of a graph $G$ is an {\em \extB of length $2k+1$, ($k>1$)} 
whenever we can find a partition of its vertex set into $2k+1$ subsets 
(considered as organized in a cyclic order) such that any two consecutive sets  in the list are joined by every
possible edge, while no edges are allowed between two non consecutive sets, and such that these sets are maximal for these properties. 

Observe that a \extB, as defined above is merely an odd hole expansion 
and that an \extB of length $5$ is precisely as defined in \cite{FouGiaMaiThu1995}. Moreover, a \extB as well as its complement are connected graphs.

A graph $G$ will be said a {\em minimal counter example to Conjecture \ref{Conjecture:Reed1998}} whenever $\chi(G)>\lceil\frac{\omega(G)+\Delta(G)+1}{2}\rceil$ 
and when Conjecture \ref{Conjecture:Reed1998} holds for any subgraph of $G$.

\subsection{Technical lemmas}

\begin{lem}\label{Lemma:No_X_Y_minimum_cexemple}
Let $G$ be a minimal counter example to Conjecture \ref{Conjecture:Reed1998} (if any). Then there are no two disjoint subsets $X\subseteq V(G)$ and $Y \subseteq V(G)$  such that $N(X) \subseteq N(Y)$ 
and $\chi(G[X]) \leq \chi(G[Y])$.
\end{lem}

\begin{prf}
Let $G^{'}$ be the subgraph obtained from  $G$ by deleting $X$. Since $G^{'}$ satisfies Conjecture \ref{Conjecture:Reed1998} by hypothesis, we have
$\chi(G^{'}) \leq \lceil \frac{\omega(G^{'})+\Delta(G^{'})+1}{2}\rceil$. We can then color the vertices of $X$ by using the colors appearing in $Y$ since $\chi(G[X]) \leq \chi(G[Y])$.
Since $\omega(G) \geq \omega(G^{'})$ and $\Delta(G) \geq \Delta(G^{'})$, we have \\
$\chi(G) = \chi(G^{'}) \leq \lceil \frac{\omega(G^{'})+\Delta(G^{'})+1}{2}\rceil \leq  \lceil \frac{\omega(G)+ \Delta(G)+1}{2}\rceil$, a contradiction.
\end{prf}

\begin{lem}\label{Lemma:Connected_component_expansion_minimum_cexemple}
Let $G=H(G_0\ldots G_{n-1})$ be an  expansion that is a minimal counter-example to Conjecture \ref{Conjecture:Reed1998} (if any). Then each component $G_{i}$ ($i\in\{0\ldots n-1\}$) is connected.
\end{lem}
\begin{prf} Without loss of generality assume that the subgraph induced by $G_0$ is not connected. Let $X$ and $Y$ be two subset of $V(G_0)$ inducing a connected component and suppose that $\chi(G[X]) \leq \chi(G[Y])$. We get immediately a contradiction with Lemma \ref{Lemma:No_X_Y_minimum_cexemple} since it can be easily checked that $N(X)=N(Y)$.
\end{prf}

\begin{lem}\label{Lemma:ReedPourSousGrapheQuiAtteintLeChromaticNumber}
 Let $H$ be an induced subgraph of some graph $G$ such that  $\chi(H)=\chi(G)$.
If $\chi(H)\leq\lceil\frac{\omega(H)+\Delta(H)+1}{2}\rceil$ then $\chi(G)\leq\lceil\frac{\omega(G)+\Delta(G)+1}{2}\rceil$.
\end{lem}

In \cite{AraKarSub2011} Aravind et al consider $k$-critical graphs in order to prove that every vertex in a minimum counter example to Conjecture \ref{Conjecture:Reed1998} belongs to an odd hole.

A graph $G$ is said to be {\em k-critical} if $\chi(G)=k$ and $\chi(G-v)<k$ for all $v\in V(G)$.

\begin{thm}\cite{AraKarSub2011}\label{thm:Odd_Hole_k-critical}
If $G$ is k-critical and $k > \frac{d(v)+\omega(v)+1}{2}$, for $v\in V(G)$, then $v$ must belong to some odd hole in $G$.
\end{thm}

We can extend the result of \cite{AraKarSub2011} to minimal counter examples to Conjecture \ref{Conjecture:Reed1998}.
\begin{lem} \label{Lemme:Hole} 
If $G$ is a minimal counter example to Conjecture \ref{Conjecture:Reed1998} then any vertex is contained in an odd hole.
\end{lem}
\begin{prf}
 Since $G$ is a minimal counter example to Conjecture \ref{Conjecture:Reed1998}, it follows that $G$ is k-critical. Then, for every $v\in V(G)$, $k > \frac{d(v)+\omega(v)+1}{2}$, 
and hence by Theorem \ref{thm:Odd_Hole_k-critical}, $v$ is part of some odd hole in $G$.
\end{prf}

\section{On \Wh graphs.}

A hole in a graph $G$ will be said {\em \Wh}, if any vertex of $G$ which is partial to $C$ is connected to precisely three consecutive vertices of $C$ or to  precisely two vertices at distance two on $C$. The graph $G$ itself 
will be said {\em \Wh} when all odd holes of $G$ are \Wh.

Observe that the vertices of a \wh cycle $C$ together the vertices which are partial to $C$ induce a \extB which, by construction, is not distinguished from the outside.

Lemma \ref{lem:P5etLeurComplémentsSontStroumpfs} below comes from a result already stated in \cite{FouGiaMaiThu1995}.
\begin{lem}\label{lem:P5etLeurComplémentsSontStroumpfs}
If $G$ is a ($P_5,\overline{P_5}$)-free graph then $G$ is \wh.
\end{lem}
\begin{prf}
Let $C$ be some odd hole in $G$ and $x$ be a vertex partial to $C$. Since $G$ is $P_5$-free, $C$ has length $5$.

The neighbours of $x$ in $C$ are either two independant vertices or three consecutive vertices, otherwise the vertices of $C$ together with $x$ would contain an induced $P_5$ or $\overline{P_5}$, a contradiction.
\end{prf}

\begin{thm} \label{Theorem:Structure_Extended_bouées} 
Let $G$ be a \wh graph. Any two distinct \extBs are vertex disjoint or one is contained in the other.
\end{thm}
\begin{prf}
Let $\mathcal{B}_1$ and $\mathcal{B}_2$ be two distinct \extBs of $G$ such that \\
$V(\mathcal{B}_1)\cap V(\mathcal{B}_2)\neq\emptyset$, $V(\mathcal{B}_1)-V(\mathcal{B}_2)\neq\emptyset$ and $V(\mathcal{B}_2)-V(\mathcal{B}_1)\neq\emptyset$. 

Observe that the vertices of $V(\mathcal{B}_1)-V(\mathcal{B}_2)$ 
as well as the vertices of $V(\mathcal{B}_2)-V(\mathcal{B}_1)$ are not partial with respect to $V(\mathcal{B}_1)\cap V(\mathcal{B}_2)$.

There is a vertex, say $x$, in $V(\mathcal{B}_1)-V(\mathcal{B}_2)$ that is connected to some vertex of $V(\mathcal{B}_1)\cap V(\mathcal{B}_2)$, otherwise $\mathcal{B}_1$ would be disconnected, a contradiction.
By the definition of a\Enleve{n} \extB, $x$ must be adjacent to all vertices of $V(\mathcal{B}_2)$.
Consequently, there must be a vertex in $V(\mathcal{B}_2)-V(\mathcal{B}_1)$, say $y$, which is adjacent to $x$.

Let $z\in V(\mathcal{B}_1)-V(\mathcal{B}_2)$ be a vertex not connected to some vertex of  $V(\mathcal{B}_1)\cap V(\mathcal{B}_2)$, then $z$ cannot be connected to $y$ since $y\in V(\mathcal{B}_2)$. 
But now, $y$ is connected to $x$ and not to $z$, another contradiction.

Consequently, all vertices in $V(\mathcal{B}_1)-V(\mathcal{B}_2)$ are adjacent to all vertices in $V(\mathcal{B}_1)\cap\mathcal{B}_2)$, in other words $\overline{\mathcal{B}_1}$ is not connected, a final contradiction.
\end{prf}

By Lemma \ref{Lemme:Hole}, every vertex in a mimimal counter example to Conjecture \ref{Conjecture:Reed1998} belongs to an odd hole, consequently :
\begin{cor}\label{cor:PartitionStroumpfGraphEnExtendedBuoys}
 Let $G$ be a \wh graph  which is a minimal counter example to Conjecture \ref{Conjecture:Reed1998}. There is a partition of the vertices of $G$ in \extBs.
\end{cor}

In \cite{FouGiaMaiThu1995} the following theorem was proved for the $(P_{5},\overline{P_{5}})$-free graphs. This result can be easily extended to \wh graphs. We give here the proof for sake of completeness.
\begin{thm}\label{Theorem:Transversal_C5} Let $G$ be a \wh graph. If $W$ is a minimum transversal of the odd cycles of $G$ then $\omega(G[W]) \leq \omega(G)-1$.
\end{thm}
\begin{prf}
 For every vertex $x$ of $W$, there exists an odd hole, denoted $C_x$, such that $W\cap V(C_x)=\{x\}$. Since $W$ is a minimal transversal of the odd holes of $G$, we call $C_x$ the private odd hole of $x$.

We have $\omega(G[W])\leq \omega[G)$. Assume that $\omega(G(W])=\omega(G)$ and let $Q$ be a maximum clique of $G[W]$.

Let $x$ be a vertex of $Q$ such that the \extB which contains $C_x$, say $\mathcal{B}(C_x)$ is minimal among all \extBs generated by private odd holes of vertices of $Q$, that is $\mathcal{B}(C_x)$ 
does not contain as a proper subset
any other $\mathcal{B}(C_y)$ with $y\in Q$.

Assume that $C_x$ has length $2k+1$ ($k>1$). We write $\mathcal{B}(C_x)=C_x(A_0,A_1\ldots A_{2k})$ since $\mathcal{B}(C_x)$ is an odd hole expansion of length $2k+1$ and we suppose that $x\in A_0$.

If $Q$ meets neither $A_1$ nor $A_{2k}$ then $Q\subseteq A_0\cup (N(\mathcal{B}(C_x))-\mathcal{B}(C_x))$. Let $y$ be a vertex of $A_1$, $\{y\}\cup Q$ is a clique of $G$, a contradiction.

We suppose now, without loss of generality, that $Q$ meets $A_1$. Let $z\in Q\cap A_1$. By minimality of $\mathcal{B}(C_x)$  and by Theorem \ref{Theorem:Structure_Extended_bouées}, $\mathcal{B}(C_x)\subseteq \mathcal{B}(C_z)$. 
Moreover, by the definition of 
a \extB, we have $\mathcal{B}(C_x)\subseteq \mathcal{B}(C_z)$.

We have $A_0\subset W$ since every odd hole obtained from $C_x$ by substituting another vertex of $A_0$ to $x$ must intersect $W$. But $C_z$ must instersect $A_0$ and $W\cap C_z\neq\{z\}$, a contradiction.
\end{prf}

Using the Strong Perfect Graph Theorem \cite{ChuRobSeyTho2006}, this result leads to

 \begin{thm} \label{Theorem:Gfree_ChiBound}
If $G$ is a $(P_6,\overline{P_6})$-free \wh graph then $\chi(G) \leq \frac{\omega(G)(\omega(G)-1)}{2}$.
 \end{thm}
\begin{prf}
Since $G$ is ($P_6,\overline{P_6}$)-free, the odd holes of $G$ have length $5$. If we remove a transversal $W$ of the $C_5$'s, we obtain a perfect graph . The perfection of $G-W$ implies that $\chi(G-W)\leq \omega(G)$ 
and by Theorem \ref{Theorem:Transversal_C5}, $\omega(G[W])\leq \omega(G)-1$.

Applying recursively this observation we get $\chi(G) \leq \frac{\omega(G)(\omega(G)-1)}{2}$.
\end{prf}

It follows from a result of King \cite{Kin2010} that if $G$ is a minimum counter-example to Conjecture \ref{Conjecture:Reed1998}
then $\omega(G) \leq \frac{2}{3}(\Delta(G)+1)$. Hence, if we restrict ourself to \wh graphs which are ($P_6,\overline{P_6}$)-free,
a minimum counter-example to this conjecture is such that $1+\sqrt{\Delta(G)+2} \leq \omega(G) \leq \frac{2}{3}(\Delta(G)+1)$.

An {\em independent buoy} is a buoy such that any set of the associated partition is a stable set.

\begin{thm} \label{Theorem:Independent_Buoy} If $G$ is a ($P_6,\overline{P_6}$)-free \wh graph where each buoy of $G$ is independent then $G$ satisfies Conjecture \ref{Conjecture:Reed1998}.
\end{thm}
\begin{prf}
By Corollary \ref{cor:PartitionStroumpfGraphEnExtendedBuoys}, there is a partition of the vertex set into \extBs.

Let $W$ be a minimum transversal of the odd holes.
We get immediately $\chi(G) \leq \chi(G-W) + \chi(G[W])$.
Moreover, since  $G-W$ and $G[W]$ does not contain any odd hole nor the complement of an odd hole, these graphs are perfect (\cite{ChuRobSeyTho2006}).

Let $G^{*}$ be the simple graph obtained from $G$ by shrinking each \extB of the partition of $G$ and deleting multiple edges.
It is an easy task to see that $2 \leq \omega(G)=\omega(G-W)=2\omega(G[W])=2\omega(G^{*})$. Hence we have $\chi(G) \leq \lceil \frac{3 \omega(G)}{2}\rceil$.

Let $v$ be a vertex contained in a maximum clique of $G$. Then $\Delta(G) \geq d(v) \geq 5(\omega(G)-1)+2$ and $\lceil \frac{\omega(G)+\Delta(G) +1}{2}\rceil \geq \lceil \frac{\omega(G)+5(\omega(G)-1)+2+1}{2}\rceil=3\omega(G)-1$.

We have thus
$\chi(G) \leq \lceil \frac{\omega(G)+\Delta(G) +1}{2}\rceil$ as soon as $\lceil \frac{3 \omega(G)}{2} \rceil \leq 3\omega(G)-1$, a contradiction.

\end{prf}

An {\em full \extB} is a \extB such that any set of the associated partition is a clique. We have immediately by Corollary \ref{Corollary:ReedsConjecturePourSpecialOddHole} 
that a full \extB satisfies Conjecture \ref{Conjecture:Reed1998}.

\begin{thm} \label{Theorem:Full_Buoy} If $G$ is a $\overline{P_6}$-free \wh graph where each \extB is full then $\chi(G) \leq \lceil \frac{3\omega(G)}{2}\rceil$.
\end{thm}

\begin{prf}
Since the buoys of $G$ are full, a buoy cannot be contained into another, thus  by Theorem \ref{Theorem:Structure_Extended_bouées} the buoys of $G$ are pairwise disjoint. 
Let $(B_i)_{1\leq i\leq k}$ be the set of buoys of $G$. Assume that the buoy $B_i$, $1\leq i\leq k$, has length $2l_i+1$, we write $B_i=C_{2l_i+1}(A^i_0,\ldots A^i_{2l_i})$. 
Without loss of generality we can consider that $A^i_0\cup A^i_1$ is a maximum clique of $B_i$ and set  $\omega_i=|A^i_0|+|A^i_1|$. Hence, we certainly have $|A^i_2|\leq\frac{\omega_i}{2}$ or $|A^i_{2l_i}|\leq\frac{\omega_i}{2}$.
For $i\in\{1,\ldots k\}$, let $W_i\in\{A^i_2,\ldots A^i_{2l_i}\}$ be a set of minimum size and let $W=\cup_{i=1}^{k}W_{i}$.

Since $G-W$ and $G[W]$ do not contain any odd hole nor its complement ($G$ is $\overline{P_6}$-free), theses graphs are perfect
and $\chi(G)\leq\omega(G-W)+\omega(G([W])$. Without loss of generality we can write a maximum clique of $G([W])$ as the set $\cup_{i=1}^{q} W_{i}$ for some $q$. By Theorem \ref{Theorem:Structure_Extended_bouées} 
this maximum clique of $G([W])$ leads to a clique of $G-W$ which is $\cup_{i=1}^{q} A^i_0 \cup A^i_1$. Hence $\omega(G[W])=\sum_{i=1}^{q} |W_{i}| \leq \sum_{i=1}^{q} \frac{\omega_{i}}{2} \leq \frac{\omega(G-W)}{2}$.

That is $\chi(G) \leq \frac{3\omega(G-W)}{2} \leq \frac{3\omega(G)}{2} $.

\end{prf}

\section{Applications}

We do not know in general whether a \wh graph  satisfies Conjecture \ref{Conjecture:Reed1998}. 
We are concerned here with various families of \wh graphs.

\begin{thm} \label{Theorem:general} If $G$ is a $P_6$-free \wh graph then $G$ satisfies Conjecture \ref{Conjecture:Reed1998} 
or $G$ contains a subgraph isomorphic to a $P_{4}(C_{5},C_{5},C_{5},C_{5})$ and a subgraph isomorphic to $C_3(C_5,C_5,C_5)$.
\end{thm}
\begin{prf}

Suppose that $G$ is  a $P_6$-free \wh graph being a minimal counter example to Conjecture \ref{Conjecture:Reed1998}.
We can consider that $G$ is connected. Since the graph is $P_6$-free, the odd holes of $G$ have length $5$. 

By Corollary \ref{cor:PartitionStroumpfGraphEnExtendedBuoys}, there is a partition of the vertex set of $G$ into \extBs. 

Let $G^{*}$ be the graph obtained from $G$ by shrinking each buoy of the above partition in a single vertex. Observe that $G^*$ is $C_5$-free.

If $G^{*}$ has only one vertex then $G$ is a $C_{5}$ expansion and the result follows from Theorem \ref{Theorem:C5_Expansion}.

Assume that $G^{*}$ contains an induced path on four vertices $B_{1}B_{2}B_{3}B_{4}$. Since each \extB of $G$ contains an induced $C_{5}$, this $P_{4}$ leads to a subgraph isomorphic to the expansion
$P_{4}(C_{5},C_{5},C_{5},C_{5})$ as a subgraph of $G$. 

If $G^{*}$ is $P_{4}$-free and contains at least two vertices, it is well known (see Seinsche \cite{Sei1974}) that its complement
is not connected.  Henceforth, $\overline{G}$ itself is not connected and $G$ satisfies Conjecture \ref{Conjecture:Reed1998} by Theorem \ref{Theorem:Complement}.

Moreover, by Theorem \ref{Theorem:ReedsForBipartiteExpansion} we can suppose that $G$ is not bipartite. Consequently $G^*$ contains a triangle, that means that $G$ contains a subgraph isomorphic to $C_3(C_5,C_5,C_5)$.
\end{prf}

Theorem \ref{Theorem:general} above implies that any $P_6$-free \wh graph   not containing some fixed subgraph of the expansion 
$P_{4}(C_{5},C_5,C_5,C_5)$ nor some subgraph of the expansion of $C_3(C_5, C_5, C_5)$ satisfies Conjecture \ref{Conjecture:Reed1998}.

For example, Conjecture \ref{Conjecture:Reed1998} holds for $P_6$-free \wh graphs of $\mathcal{G}$ with no induced $K_6$, since $K_6$ is a subgraph of $C_3(C_5,C_5,C_5)$.

Moreover, by this way we get shorter proofs of results given in \cite{AraKarSub2011}.

\begin{cor} \cite{AraKarSub2011} Any $(C_{4},P_{5})$-free graph satisfies Conjecture \ref{Conjecture:Reed1998}.
\end{cor}
\begin{prf} Let $G$ be a ($C_4,P_5$)-free graph. Since $G$ is $P_5$-free, the odd holes of $G$ have length $5$. It is not difficult to check that a vertex partial to some odd hole of $G$, say $C$, 
is precisely connected to $3$ consecutive vertices of $C$.
 By definition, a $(C_{4},P_{5})$-free graph is  a 
$P_6$-free \wh graph. Since a $P_{4}(C_{5},C_{5},C_{5},C_{5})$  contains a $C_{4}$, the result follows from Theorem \ref{Theorem:general}
\end{prf}

\begin{cor}  Any $(P_{5},\overline{P_{5}},Dart)$-free graph satisfies Conjecture \ref{Conjecture:Reed1998}
\end{cor}
\begin{prf} By Lemma \ref{lem:P5etLeurComplémentsSontStroumpfs}, a  $(P_{5},\overline{P_{5}})$-free graph is \wh. 
Moreover, it is obviously a $P_6$-free graph. Since a $P_{4}(C_{5},C_{5},C_{5},C_{5})$  contains a $Dart$, the result follows from Theorem \ref{Theorem:general}.
\end{prf}

\begin{cor} \cite{AraKarSub2011} Any $(P_{5},\overline{P_{5}}, Dart, \overline{P_{2} \cup P_{3}})$-free graph satisfies Conjecture \ref{Conjecture:Reed1998}
\end{cor}

\begin{cor}  Any $(P_{5},Kite)$-free graph satisfies Conjecture \ref{Conjecture:Reed1998}
\end{cor}

\begin{prf} 
Let $G$ be a ($P_5, Kite$)-free graph. Since $G$ is $P_5$-free, the odd holes of $G$ have length $5$. It is not difficult to check that a vertex partial to some odd hole of $G$, say $C$, 
is precisely connected to $2$ vertices at distance $2$ on $C$.
By definition $G$ is  \wh. Moreover $G$ is $P_6$-free. Since a $P_{4}(C_{5},C_{5},C_{5},C_{5})$  contains a $Kite$, 
the result follows from Theorem \ref{Theorem:general}
\end{prf}

\begin{cor}  \cite{AraKarSub2011} Any $(P_{5},Kite, Bull, (K_{3} \cup K_{1})+K_{1})$-free graph satisfies Conjecture \ref{Conjecture:Reed1998}
\end{cor}

\subsection{$(Chair, Bull)$-free graphs}
$\;$\\

\begin{lem}\label{lemma:Chair-bull-free}
Let $G$ be a ($Chair, Bull$)-free graph $G$ and $C_{2k+1}$ ($k>1$) be an odd hole of $G$. Let $x$ be a vertex of $G$ partial to $C_{2k+1}$.\\
One of the following holds~:
\begin{enumerate}
\item $x$ is adjacent to precisely $3$ consecutive vertices of $C_{2k+1}$,
\item  $k=2$ and $x$ is  adjacent to precisely four vertices of $C_{2k+1}$.
\end{enumerate}
\end{lem}
\begin{prf}
Let us write $C_{2k+1}=v_0v_1\ldots v_{2k}$. Without loss of generality we can assume that $x$ is adjacent to $v_0$ and not adjacent to $v_{2k}$.

The vertex $x$ must have at least one neighbour in $\{v_1,v_2\}$ otherwise the set $\{x,v_{2k},v_0,v_1,v_2\}$ would induce a $Chair$, a contradiction.
If $x$ is connected to $v_1$ and not to $v_2$, the set $\{x,v_{2k},v_0,v_1,v_2\}$ would induce a $Bull$, a contradiction.

If $x$ is connected to $v_2$ but not to $v_1$, the vertex $x$ must be adjacent $v_{2k-1}$ or the vertices $v_{2k-1}$, $v_{2k}$, $x$, $v_0$ and $v_1$ would induce a $Chair$, a contradiction. 
Consequently, $k=2$, otherwise the vertices $v_2$ and $v_{2k-2}$ are distinct and independent and $\{v_{2k-2}, v_{2k-1}, v_{2k}, x, v_2\}$ induces a $Bull$ when $x$ is adjacent to $v_{2k-2}$ and a $Chair$ otherwise. 
But now the vertex $x$ together with $v_1$, $v_2$, $v_3$ and $v_4$ would induce a $Bull$, a contradiction.

It follows that $x$ is adjacent to $v_1$ and $v_2$.

If $x$ has another neighbour on $C_{2k+1}$, say $y$, we have $y=v_{2k-1}$, otherwise the vertices $x$, $y$, $v_0$, $v_1$, $v_{2k}$ iduce a $Bull$, a contradiction. Once again, we have $k=2$, or the vertices  
$v_2$ and $v_{2k-2}$ being distinct and independent, the set $\{v_{2k-2}, v_{2k-1}, v_{2k},x, v_0, v_2\}$ would contain an induced $Chair$ when $x$ and $v_{2k-1}$ are not adjacent and an induced $Bull$ otherwise, a contradiction.

Hence, $k=2$ and $x$ is adjacent to precisely four vertices of the cycle $C_5$.
\end{prf}

Let us denote $\mathcal{F}$ the following set of graphs $\{House,Kite, Gem, C_5\}$ (see Figure~\ref{fig:ChairHouseBullDartKite}).

 \begin{thm} \label{Theorem:Bull_House_Chair}  If $G$ is a $(Chair, Bull, F)$-free graph with $F\in\mathcal{F}$, then $G$ satisfies Conjecture \ref{Conjecture:Reed1998}.
\end{thm}
\begin{prf}
Let $G$ be a ($Chair, Bull, F$)-free graph. Assume that $G$ is a minimal counter example to Conjecture \ref{Conjecture:Reed1998}.
We can consider that $G$ is connected. 
Since $G$ is $F$-free, by Lemma \ref{lemma:Chair-bull-free}, $G$ is a \wh graph. Since $G$ is $Chair$-free, it is not difficult to check that the \extBs are full.
By Corollary \ref{cor:PartitionStroumpfGraphEnExtendedBuoys} there is a partition of the vertex set of $G$ into \extBs.

Let $G^{*}$ be the graph obtained from $G$ by shrinking each \extB of the above partition in a single vertex. Observe that $G^*$ is  odd hole free.

If $G^{*}$ has only one vertex then $G$ itself is a full odd hole expansion. By Corollary \ref{Corollary:ReedsConjecturePourSpecialOddHole}, Conjecture \ref{Conjecture:Reed1998} holds for $G$.

In addition, $G^{*}$ is $P_{4}$-free. As a matter of fact, since each vertex of $G^{*}$ represents an odd hole, such a $P_4$ in $G^{*}$ would represent a subgraph of $G$ which is not $Chair$-free, a contradiction.

Consequently, if $G^{*}$ contains at least two vertices, it is well known (see Seinsche \cite{Sei1974}) that its complement
is not connected.  Henceforth, $\overline{G}$ itself is not connected and $G$ satisfies Conjecture \ref{Conjecture:Reed1998} by Theorem \ref{Theorem:Complement}.
\end{prf}

\begin{cor}  \cite{AraKarSub2011} Any $(Chair, \overline{P_{5}}, Bull, K_{1}+C_{4})$-free graph satisfies Conjecture \ref{Conjecture:Reed1998}
\end{cor}

\bibliographystyle{plain}
\bibliography{BibliographieReed}

\begin{thebibliography}{1}

\bibitem{AraKarSub2011}
N.R. Aravind, T.~Karthick, and C.R. Subramanian.
\newblock Bounding $\chi$ in terms of $\omega$ and {$\Delta$} for some classes
  of graphs.
\newblock {\em Discrete Mathematics}, 311:911--920, 2011.

\bibitem{BonMur08}
J.A. Bondy and U.S.R. Murty.
\newblock {\em Graph Theory}, volume 244 of {\em Graduate Text in Mathematics}.
\newblock Springer, 2008.

\bibitem{ChuRobSeyTho2006}
M.~Chudnovsky, N.~Robertson, P.D. Seymour, and R.~Thomas.
\newblock The {S}trong {P}erfect {G}raph {T}heorem.
\newblock {\em Annals of Math.}, 164:51--229, 2006.

\bibitem{FouGiaMaiThu1995}
J.L. Fouquet, V.~Giakoumakis, F.~Maire, and H.~Thuillier.
\newblock On graphs without ${P}_{5}$ and $\overline{P_{5}}$.
\newblock {\em Discrete Mathematics}, 146(33-44), 1995.

\bibitem{FouVan2011}
J.L. Fouquet and J.M. Vanherpe.
\newblock Reed's conjecture on hole expansions.
\newblock Technical report, L.I.F.O., 2011.

\bibitem{Kin2010}
A.D. King.
\newblock Hitting all maximum cliques with a stable set using lopsided incident
  transversals.
\newblock {\em Journal of Graph Theory}, 2:111, 2010.

\bibitem{Rab2008}
L.~Rabern.
\newblock A note on {R}eed's conjecture.
\newblock {\em SIAM Journal on Discrete Mathematics}, 22:820--827, 2008.

\bibitem{Ree1998}
B.~Reed.
\newblock $\omega$, ${\Delta}$ and $\chi$.
\newblock {\em Journal of Graph Theory}, 27:177--212, 1998.

\bibitem{Sei1974}
D.~Seinsche.
\newblock On a property of the class of $n$-colorable graphs.
\newblock {\em Journal of Combinatorial Theory}, Series B(16):191--193, 1974.

\end{thebibliography}

\end{document}